\documentclass[twocolumn,preprintnumbers,amssymb,amsmath,superscriptaddress,nofootinbib]{revtex4}
\usepackage[dvips]{graphicx}
\usepackage{dcolumn}
\usepackage{bm}
\usepackage{epsfig,amsmath,amssymb,verbatim,mathrsfs,array,layout,textcomp,amssymb,latexsym}
\input epsf.tex
\newcommand{\beq}{\begin{eqnarray}}
\newcommand{\eeq}{\end{eqnarray}}

\def\ltap{\ \raise.3ex\hbox{$<$\kern-.75em\lower1ex\hbox{$\sim$}}\ }
\def\gtap{\ \raise.3ex\hbox{$>$\kern-.75em\lower1ex\hbox{$\sim$}}\ }

\def\be{\begin{equation}}
\def\ee{\end{equation}}
\def\bea{\begin{eqnarray}}
\def\eea{\end{eqnarray}}
\newcommand{\eref}[1]{(\ref{#1})}
\newcommand{\Eref}[1]{Eq.~(\ref{#1})}

\begin{document} 

\title{
Higgs Up-Down CP Asymmetry at the LHC
}
\author{C\'edric Delaunay}
\affiliation{CERN Physics Department, Theory Division, CH-1211 Geneva 23, Switzerland}
\author{Gilad Perez}
\affiliation{CERN Physics Department, Theory Division, CH-1211 Geneva 23, Switzerland}
\affiliation{Department of Particle Physics, Weizmann Institute of Science, Rehovot 76100, Israel}
\author{Hiroshi de Sandes}
\affiliation{CERN Physics Department, Theory Division, CH-1211 Geneva 23, Switzerland}
\author{Witold Skiba}
\affiliation{CERN Physics Department, Theory Division, CH-1211 Geneva 23, Switzerland}
\affiliation{Department of Physics, Yale University, New Haven, CT 06520, USA}
\preprint{\scriptsize CERN-PH-TH/2013-202\vspace*{.1cm}\\}
\vskip .05in

\begin{abstract}
\vskip .05in

We propose a new observable designed to probe CP-violating coupling of the Higgs boson to $W$ bosons using associated Higgs production.  We define an asymmetry that measures the number of leptons from $W$ decays relative to the plane defined by the beam line and the Higgs boson momentum. The orientation of that plane is determined by the direction of fermions in the initial state, so that in a proton-proton collider it requires rapidity cuts that preferentially select quarks over antiquarks. 

\end{abstract}

\maketitle

\section{Introduction}

The recent discovery  of a Higgs-like particle at the LHC~\cite{ATLASdisco,CMSdisco} launched a program of detailed studies of properties of the $\sim 126$~GeV boson. The initial measurements indicate that its couplings are close to the ones predicted by the Standard Model (SM)~\cite{Higgsfits}. With time, either credible deviations from the SM properties will emerge or the SM Higgs boson will be confirmed within shrinking experimental errors. This is analogous to the program of precision electroweak measurements, which searched for deviations from the SM among numerous properties of the electroweak gauge bosons and four-fermion interactions. Investigating Higgs couplings opens up sensitivity to new physics that couples to the Higgs boson that previous measurements were unable to probe directly. 

Naturalness arguments suggest existence of new states with couplings to the Higgs boson. If naturalness is indeed a useful guide, new physics is related to the top quark and massive gauge bosons and it is likely to be noticeable in the Higgs couplings to these particles. Higgs couplings to gluons and photons are known to be sensitive probes of new physics~\cite{ManoharWise} because these couplings are generated at loop level in the SM leaving room for relatively large contributions from new physics. Within the SM the Higgs couplings to the massless gauge bosons are dominated by the top quark and $W$ loops in the SM and therefore indirectly probe the Higgs couplings to these particles. However, it would be desirable to be able to probe the modifications to the Higgs coupling in a more direct way.  
As we will discuss shortly, the most general Higgs couplings to the massive gauge bosons can involve several Lorentz structures, with different CP properties. It would be very interesting to pin down these couplings as accurately as possible. CP violation in the quark sector is consistent with the single CP-violating phase in the CKM matrix. The bounds on CP violation in the light quark and lepton sectors are very stringent relegating new CP-violating physics to very high scales. It is an intriguing possibility that CP-violating interactions involving the Higgs occur at lower scales that could be accessible at the LHC\@.

We focus here on Higgs couplings to $V=W,Z$ weak bosons and we write the most generic $hV_\mu V_\nu$ vertex as 
\beq
\label{hVV}
- ig_Vm_V\left[A_V \eta_{\mu\nu} +B_V p_{1\nu}p_{2\mu} + C_V \epsilon_{\mu\nu\alpha\beta}p_1^\beta p_2^\alpha\right]\,, 
\eeq
where $p_{1,2}$  are the incoming four-momenta of $V_\mu$ and $V_\nu$ and $\epsilon^{0123}=1$. We factored out the couplings $g_W=g$, $g_Z=\sqrt{g^2+g^{\prime\,2}}$, where $g$ and $g'$ are the $SU(2)_L\times U(1)_Y$ gauge couplings, respectively. Meanwhile, $A,B,C$ are form factors that are functions of the Lorentz scalars $p_{1,2}^2$ and $p_1\cdot p_2$. (See Ref.~\cite{assocprod} for a recent discussion of the role of these form factors in associated Higgs production.) The SM predicts at tree level $A_V^{\rm SM}=1$ and $B_V^{\rm SM}=C_V^{\rm SM}=0$. The first two couplings in Eq.~\eqref{hVV} are CP-conserving while, in the presence of them, the third one violates CP. Note that in the $A_V=B_V=0$ limit, there is a parity assignment, where $h$ is parity odd, so that CP is conserved. Such an assignment no longer exists when $C_V$ is present together with either $A_V$ or $B_V$, leading to CP-violating effects coming from the interference between $A_V,B_V$ and $C_V$ couplings. 

There is already some evidence regarding the CP properties of the $126$~GeV boson based on its decays into $Z$ pairs using angular analysis of the 4-lepton channel~\cite{CMS_4leptons, ATLAS_4l}. The decays are consistent with CP-even couplings indicating that either the Higgs boson is a CP eigenstate and CP-violating couplings to the $Z$'s are sub-dominant, or that the scalar is a mixture of CP-even and CP-odd states with a larger CP-even component.  Even though the $C_Z$ form factor is already constrained we will argue in Section~\ref{sec:eft} that couplings to the $W$ boson pairs need not follow exactly the same pattern as the couplings to the $Z$ pairs as these couplings could arise from several independent higher-dimensional operators. 

Constraining $C_W$ in the $hWW$ vertex through Higgs decays seems more difficult compared to probing $C_Z$ in $h\to 4 l$. The challenge stems from missing energy in the $h\to 2l2\nu$ channel, and missing energy and poor jet resolution in the $h\to l\nu q \bar{q'}$ channel. Measurements of the total decay rate $h\to WW^*$ are only sensitive to the square of $C_W$ and furthermore it is not possible to disentangle non-SM values of $A_W$ and $B_W$ from $C_W\neq 0$. 

We argue in this letter that the associated $Wh$ production channel offers a complimentary probe of the presence of the CP-odd interaction in the $hWW$ vertex. A key difference between the two channels is that the $h\to WW^*$ decay is only quadratically sensitive to $B_W$ and $C_W$ coefficient evaluated at $(p_1+p_2)^2=m_h^2$, whereas in $Wh$ production the momentum transfer is controlled differentially by varying the $Wh$ invariant mass. We propose a new observable related to the triple product  $\vec l\cdot (\vec h\times \vec q)$, where $\vec l$, $\vec h$ and $\vec q$ are the three-momenta of the charged lepton from the $W$ decay, the Higgs boson, and the initial quark in the $q\bar q'$ ($q'\neq q$) partonic collision, respectively. Since the triple-product is a Lorentz pseudo-scalar, the proposed asymmetry is induced by the interference between CP-conserving and CP-violating couplings and its magnitude is linearly proportional to $C_W$. 

Other observables were proposed to reveal the presence of CP-odd Higgs interactions in associated $Wh$  production~\cite{Godbole}. The latter are also sensitive to CP-even interactions and their measurements  are thus complimentary to the one proposed in this work. CP violation in the $hWW$ vertex could also be revealed in Higgs production through vector boson fusion~\cite{Zeppenfeld}, although extracting the  $W$ contribution from the $Z$ may be an obstacle in this channel.

The reason for defining an observable proportional to a triple product is easy to understand as the $C_V$ vertex in \eref{hVV} contains the antisymmetric $\epsilon$ tensor. An observable sensitive to $C_V$ must therefore rely on measuring three linearly independent three-vectors. Let us first consider the $h\to ZZ^* \to 4 l$ process in the Higgs rest frame. A suitable observable must be then constructed out of the momenta of three of the leptons, as the fourth one is restricted by momentum conservation. In practice, a triple product sensitive to $C_Z$ is proportional to the angle between planes defined by the lepton pairs in the Higgs rest frame~\cite{Gao}. In the associated production partonic process $q \bar{q} \to W h$, the three linearly independent vectors are chosen to be the beam direction, the Higgs momentum and the $W$ polarization. As we demonstrate later, there is actually no need to determine the polarization of the $W$. Instead one could rely on measuring the momentum of the lepton created in the $W\to l \nu$ decay. Measuring the lepton momentum is straightforward experimentally and it turns out to be a good substitute for $W$ polarization. 

The outline of the paper is as follows. In the next section we compute helicity amplitudes for the parton-level process $q \bar{q'} \to W h$.  In Section~\ref{sec:asymmetry} we define our asymmetry observable and present sensitivity estimates for the 14~TeV run of the LHC\@.  In Section~\ref{sec:eft} we comment on the possible origin of non-SM couplings $B_V$ and $C_V$ in an effective field theory. We also discuss direct and indirect bounds on the coefficients of effective operators that can lead to non zero $C_V$. We conclude in Section~\ref{sec:conclusions}.

\section{Helicity amplitudes for $Wh$ production}
\label{sec:helicity}

We start by evaluating the cross-section for the partonic process $q\bar{q}^{\,\prime}\to Wh$ using the generic $hWW$ vertex in \Eref{hVV}. We assume that $W$ and $h$ are produced on shell and rely on the narrow width approximation (NWA) to subsequently include the $W\to l\nu$ decay. We assume that Higgs decays to $b\bar{b}$ because it is the channel with the largest branching ratio, but the particular decay channel is not important for our result. Higgs decay products do not carry any important information about the interaction in \Eref{hVV} that we want to probe because Higgs is a scalar. Higgs decay products are crucial only for Higgs identification and determination of its momentum. 

Consider first the partonic process $u\bar d \to W^+ h$ (and an analogous calculation for $d\bar u \to W^- h$) with on-shell Higgs and $W$ boson. Neglecting quark masses, the helicities of the initial quarks are fixed by the $V-A$ nature of the $W$ interaction. Using the $hWW$ vertex in Eq.~\eqref{hVV}, one finds the following amplitudes~\footnote{In a frame where the $W$ momentum reads $(q_0,0,0,q)$, the $W$ polarization four-vectors are $\varepsilon_\pm^\mu = (0,1,\pm i,0)/\sqrt{2}$ and $\varepsilon_0^\mu = (q,0,0,q_0)/m_W$ for $\lambda=\pm 1$ and $\lambda=0$ helicities, respectively.} 
\begin{gather}
\mathcal{M}_\pm^p =\pm gm_W \mathcal{A}_T\frac{\left(1\mp \cos\theta\right)}{\sqrt{2}}e^{\pm i \gamma}\,,\label{Mp1}\\
 \mathcal{M}_0^p = -gm_W\mathcal{A}_L \sin\theta\,,\label{Mp2}
\end{gather}
for producing transverse $W$ of helicity $\lambda=\pm 1$ or longitudinal $W$ of helicity $\lambda=0$ in the final state. In the formulas above, $\theta$ is the  scattering angle in the center-of-mass frame (cmf), while $\sqrt{\hat s}$ is the cmf energy and $\beta\equiv \sqrt{1-4m^2/\hat s+\delta^2}$, with $m^2\equiv (m_W^2+m_h^2)/2$ and $\delta\equiv (m_h^2-m_W^2)/\hat s$.  Finally, the proportionality factors are given by
$\mathcal{A}_T = \sqrt{A_W^2+(C_W\hat s\beta)^2/4}$, 
$\mathcal{A}_L = A_W(1-\delta)+B_W\hat s\beta^2/2$
and
\beq
\tan\gamma = \frac{C_W\hat s \beta}{2A_W}\,.
\eeq

It is worth noting that $\tan \gamma$ encodes information about the CP-violating part of the $hWW$ interaction and is proportional to $C_W$. At the kinematic threshold for $Wh$ production $\beta=0$ and hence $\gamma=0$. At threshold, Higgs momentum vanishes and therefore the triple product involving Higgs momentum and $W$ polarization vanishes as well. Likewise, the amplitude for producing longitudinally polarized $W$'s, $ \mathcal{M}_0^p$ in \Eref{Mp2}, is independent of $\gamma$ and thus insensitive to CP violation. The longitudinal polarization vector is parallel to the $W$ momentum, and hence proportional to Higgs momentum in the cmf. Consequently, the triple product vanishes in this case. 

\begin{figure}
\centering
\includegraphics[scale=0.5]{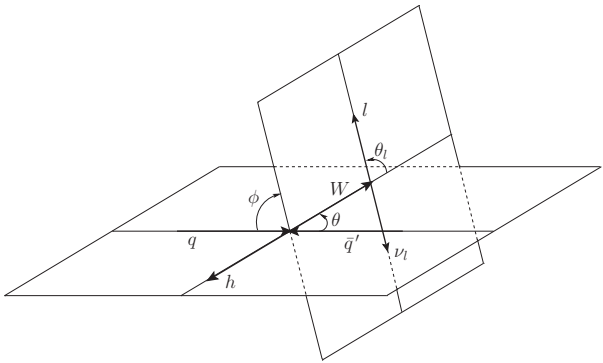}
\caption{Definition of the production and decay angles. The $W$ and $h$ directions are drawn in the $q\bar{q}^{\,\prime}$ center-of-mass frame, while the leptons are drawn in their parent $W$ rest frame. $\phi$ is the angle between the production plane and the $W$ decay plane.}
\label{figangles}
\end{figure}

The amplitudes for the subsequent decay of polarized $W^+ \to l^+\nu$    ($l=e,\mu$) are (neglecting lepton masses)
\beq\label{Md}
\mathcal{M}_\pm^d = \mp \frac{gm_W}{\sqrt{2}}\frac{\left(1\pm \cos\theta_l\right)}{\sqrt{2}}e^{\pm i\phi}\,,\ \mathcal{M}_0^d = \frac{gm_W}{\sqrt{2}}\sin\theta_l
\eeq
for the transverse ($\mathcal{M}_\pm^d$) and longitudinal ($\mathcal{M}_0^d$) bosons. As illustrated in Fig.~\ref{figangles}, $\theta_l$ is the angle in the $W$ rest frame between the charged lepton momentum and the direction of flight of the $W$ as seen from the cmf, while $\phi$ is the azimuthal angle between the production plane, defined by the momenta of the incoming quark and the outgoing Higgs boson, and the $l\nu$ decay plane in the cmf. Note that the decay amplitudes carry phases for non-zero azimuthal angles that depend on the helicity $e^{i \lambda \phi}$, where $\lambda=\pm1,0$. 

The differential cross-section for $u\bar d \to W^+ h \to l^+ \nu h$ reads $d\hat\sigma =1/(3\hat s)\left|\overline{\mathcal{M}}\right|^2d{\rm PS}_{l\nu h}\,$, where $\left|\overline{\mathcal{M}}\right|^2$ is the associated amplitude squared averaged (summed) over the initial (final) fermion spins, the factor of $1/3$  comes form  color average,
and $d{\rm PS}_{l\nu h}$ is the three-body relativistic phase-space for $l^+\nu h$ final states. Using the NWA,  the cross section is well approximated by
\beq\label{dsigma}
d\hat\sigma \simeq \frac{\pi}{12\hat s m_W\Gamma_W}\left|\sum_{\lambda} \mathcal{M}^p_{\lambda}\mathcal{M}^d_{\lambda}\right|^2  d{\rm PS}_{Wh}\ d{\rm PS}_{l\nu}\,,
\eeq 
where $\Gamma_W\ll m_W$ is the $W$ width. The amplitudes $\mathcal{M}^{p,d}_{\lambda}$ are defined in Eqs.~\eref{Mp1}, \eref{Mp2}, and~\eref{Md}, while the $W$ helicity sum runs over $\lambda=\pm1,0$. The phase space is expressed as a product of $d{\rm PS}_{Wh}$ and $d{\rm PS}_{l\nu}$, which are the two-body relativistic phase-spaces for the processes $u\bar d \to W^+h$  and $W^+\to l^+\nu$, respectively. These reduce to $d{\rm PS}_{Wh}=(\beta/16\pi) d\cos\theta$ in the cmf and  $d{\rm PS}_{l\nu}= (1/32\pi^2)d\cos\theta_l d\phi$ in the $W$ rest frame.

The absolute value square of the helicity sum in \Eref{dsigma} decomposes as
\beq
\left|\sum_{\lambda} \mathcal{M}^p_{\lambda}\mathcal{M}^d_{\lambda}\right|^2 = \sum_\lambda \left|\mathcal{M}^p_{\lambda}\right|^2\left|\mathcal{M}^d_{\lambda}\right|^2 \hspace{2,3cm}\nonumber \\ 
 +\ 2\sum_{\lambda>\lambda'}{\rm Re}\left[\mathcal{M}^p_{\lambda}\mathcal{M}^{p\,*}_{\lambda'}\mathcal{M}^d_{\lambda}\mathcal{M}^{d\,*}_{\lambda'}\right]\,,
\eeq
where the second term collects interferences between different helicity amplitudes. 
Using Eqs.~\eref{Mp1}, \eref{Mp2} and \eref{Md} it is straightforward to check that interference effects vanish when averaged over the azimuth angle $\phi$, since helicity is conserved, and that $d^2\hat\sigma/d\cos\theta d\cos\theta_l$ only depends quadratically on $C_W$. However, any observable probing the azimuthal angle distribution is linearly sensitive to $C_W$. The simplest of such observables is the up-down asymmetry
\beq\label{AhatCP} 
\hat A_{\rm CP} \equiv  \frac{\hat \sigma_{\phi>0}-\hat \sigma_{\phi<0}}{\hat \sigma_{\phi>0}+\hat \sigma_{\phi<0}}=-\frac{9\pi}{16}\sin\gamma\left(\frac{\mathcal{A}_T\mathcal{A}_L}{2\mathcal{A}_T^2+\mathcal{A}_L^2}\right)\,,
\eeq
where $\hat\sigma_{\phi<0} = \int_{-\pi}^0 d\hat\sigma/d\phi$ and $\hat\sigma_{\phi>0} = \int^{\pi}_0 d\hat\sigma/d\phi$. $\hat A_{\rm CP}$ is a measure of how often the charged lepton from the $W$ decay flies above the production plane, relative to below that plane, where above (below) the plane is defined by $\vec l\cdot (\vec h \times \vec u)>0$ $(<0)$. We describe next how to probe and what the expectations are for this asymmetry in both $p\bar p$ and $pp$ colliders.

\section{Up-down asymmetry at hadron colliders}
\label{sec:asymmetry}
Consider the hadronic process $h_1 h_2 \to W^+h \to l^+\nu b\bar b$ with $\sqrt{s}$ energy in the cmf. 
We define the asymmetry 
\beq\label{ACP}
A_{\rm CP} \equiv \frac{N_\uparrow-N_\downarrow}{N_\uparrow+N_\downarrow}\,,
\eeq
where $N_\uparrow$ $(N_\downarrow)$ is the number of events satisfying $\vec l \cdot (\vec h \times \vec h_1) >0$ $(<0)$, {\it i.e} with a charged lepton flying ``above'' (``below'') the production plane. The differential cross-section for the above process is~\footnote{A sum over all the possible $q\bar{q}^{\,\prime}$ initial states is understood.}
\beq\label{hdiff}
 \frac{d^2\sigma}{d\tau d\phi} =  \mathcal{L}_{q\bar{q}^{\,\prime}}(\tau) \frac{d\hat\sigma}{d\phi}(\tau,\phi) +\mathcal{L}_{\bar{q}^{\,\prime} q}(\tau)\frac{d\hat\sigma}{d\phi}(\tau,-\phi)\,,
\eeq
where $\tau\equiv \hat s/s$ and $\mathcal{L}_{ij}(\tau)\equiv \int_\tau^1 \frac{dx}{x} \ f_{i/h_1}(x) f_{j/h_2}(\tau/x)\,$, with $f_{i/h_a}(x)$ is the parton distribution function (PDF) controlling the probability of finding a parton $i$ with a fraction $x$ of the hadron $h_a$ momentum. The $\bar{q}^{\,\prime}q$ initial state is related to the $q\bar{q}^{\,\prime}$ one through a parity transformation under which the triple product $\vec l\cdot (\vec h \times \vec q\,)\propto \sin\phi$ flips sign, hence the extra minus sign in the second term of Eq.~\eqref{hdiff}. The number of ``upward'' events is thus 
\beq
N_{\uparrow}=\int_{\tau_0}^1d\tau\,\left[ \mathcal{L}_{q\bar{q}^{\,\prime}}(\tau)\hat\sigma_{\phi>0}(\tau)+\mathcal{L}_{\bar{q}^{\,\prime}q}(\tau)\hat\sigma_{\phi<0}(\tau)\right]\,,
\eeq
with $\tau_0=(m_W+m_h)^2/s$, while $N_{\downarrow}$, the number of ``downward'' events, is obtained from $N_{\uparrow}$ through exchanging $\hat\sigma_{\phi>0}$ and $\hat\sigma_{\phi<0}$. 

A completely analogous asymmetry can be defined for the process $h_1h_2\to W^-h\to l^-\bar \nu b\bar b$. The up-down asymmetry is expected to be of opposite sign relative to the process leading to $l^+$ because charge conjugation of the $W$ decay amplitude is equivalent to taking $\phi\to -\phi$. The statistical significance of the asymmetry for negatively charged leptons, however, would be smaller since down quark PDFs are smaller than up quark PDFs in the proton. Although the up-down asymmetry in $W^-$ associated Higgs production is less sensitive to the CP-odd vertex in $hWW$, its measurement could be used as an independent test of the asymmetry measured in $W^+$ associated production. 

We now evaluate the expected up-down asymmetry in Eq.~\eqref{ACP} at the Tevatron ($\sqrt{s}=1.96\,$TeV) and the LHC with $\sqrt{s}=14\,$TeV. For illustration, we focus on the case where $A_W=A_W^{\rm SM}=1$, $B_W=B_W^{\rm SM}=0$ and $C_W\neq0$. Although $C_W$ could be a generic form factor, we consider constant $C_W$ for simplicity. The leading contributions in an effective field theory expansion to the form factors in \Eref{hVV} are momentum independent, as we discuss in the next section. 
Hence, we take $C_W=4/\Lambda^2$, where $\Lambda$ is the scale of the dimension six operator $\widetilde{\mathcal{O}}_{WW}$ defined in \Eref{eq:d6odd}. We use the CTEQ6L1~\cite{CTEQ} PDF sets at leading order to compute the hadronic cross-sections and MadGraph 5~\cite{MG5} to simulate events.

In $p\bar p$ collisions at the Tevatron it is far more likely that $q$ arises from the proton, {\it i.e.} $\mathcal{L}_{q\bar{q}^{\,\prime}}\gg \mathcal{L}_{\bar{q}^{\,\prime}q}$, and the up-down asymmetry is well approximated by
\beq
A_{\rm CP}^{p\bar p}\simeq\frac{\int d\tau\,\mathcal{L}_{q\bar{q}^{\,\prime}}(\tau)\left[\hat\sigma_{\phi>0}(\tau)-\hat\sigma_{\phi<0}(\tau)\right]}{\int d\tau\,\mathcal{L}_{q\bar{q}^{\,\prime}}(\tau)\left[\hat\sigma_{\phi>0}(\tau)+\hat\sigma_{\phi<0}(\tau)\right]}\,.
\eeq
For $\Lambda=500\,$GeV and $1\,$TeV, the inclusive asymmetries are $A_{\rm CP}^{p\bar p}\simeq -23\%$ and $-6.3\%$, respectively. Such asymmetries are however unlikely to be observed at the Tevatron due to small statistics.

At the LHC, the initial $pp$ state is symmetric under parity, $\mathcal{L}_{q\bar{q}^{\,\prime}}= \mathcal{L}_{\bar{q}^{\,\prime}q}$, thus without further cuts $A_{\rm CP}^{pp}=0$. Any asymmetry induced in $q\bar{q}^{\,\prime}$ events is exactly compensated by $\bar{q}^{\,\prime}q$ ones. A simple way of overcoming this is by breaking the parity invariance of the initial $pp$ state by selecting events for which the partonic cmf is boosted relative to the laboratory. As the valence quark tend to carry a larger momenta fraction than the sea anti-quarks the direction of the boost is correlated with the direction of the incoming quark and can be used to define the production plane.

\begin{figure}[htb]
\centering
\begin{tabular}{c}
\includegraphics[scale=0.59]{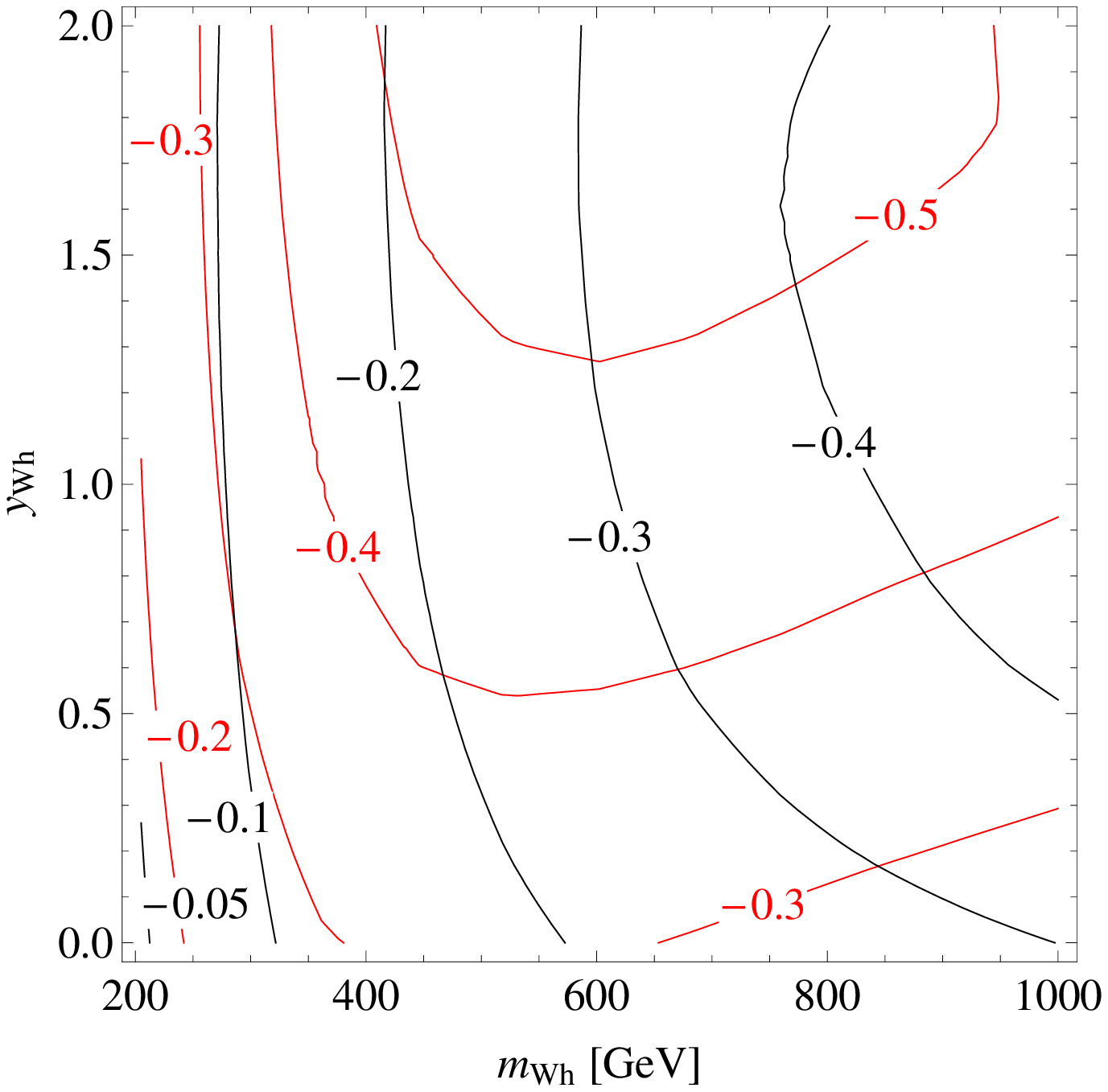}\\
\vspace{0.3cm}\\
\includegraphics[scale=0.59]{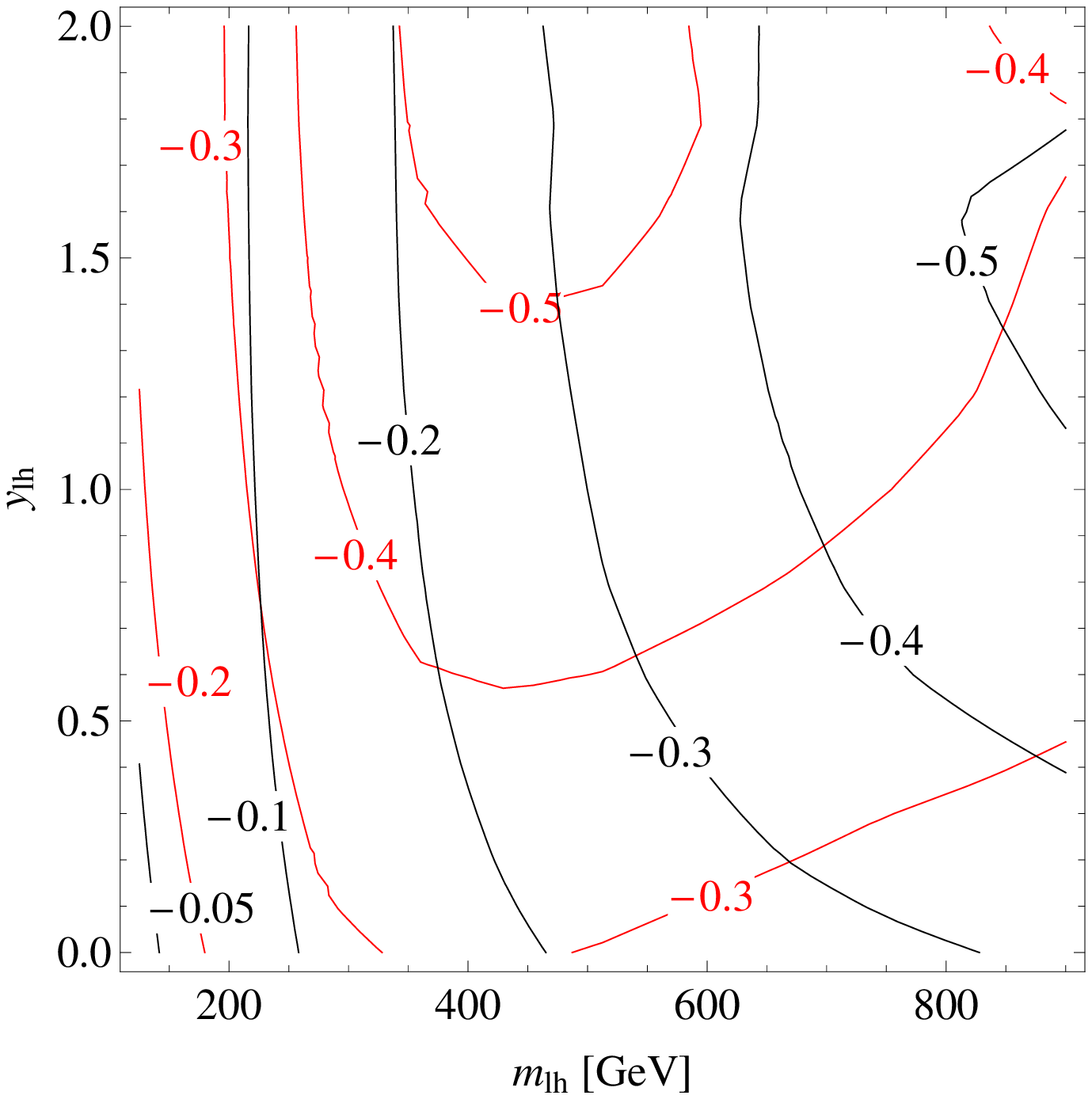}
\end{tabular}
\caption{Contours of the up-down asymmetry $A_{\rm CP}$ in associated $Wh$ production expected at the LHC with $\sqrt{s}=14\,$TeV as a function of cuts on the $Wh$ pair rapidity $y_{Wh}$ and invariant mass $m_{Wh}$  (top) and on the $lh$ pair rapidity $y_{lh}$ and invariant mass $m_{lh}$ (bottom). The $hWW$ vertex is that of Eq.~\eqref{hVV} assuming $A_W=A_W^{\rm SM}=1$, $B_W=B_W^{\rm SM}=0$ and $C_W=4/\Lambda^2$ with $\Lambda=500\,$GeV (red) and $\Lambda=1\,$TeV (black).}
\label{figApp}
\end{figure}
The boost of the partonic cmf relative to the $pp$ frame is characterized by $y_{Wh}$, the rapidity of the $W$ and $h$ bosons pair in the laboratory frame. For events with $y_{Wh}>0$ and for $\Lambda=500\,$GeV and $1\,$TeV, the resulting up-down asymmetries are $A_{\rm CP}^{pp}\simeq -14\%$ and $-4.1\%$, respectively.  The $W$ rapidity may not be reconstructed well enough experimentally due to the missing neutrino, leading to a poor estimation of the partonic cmf boost. One possible alternative is to trade the $W$ rapidity for that of the lepton and select events for which the rapidity of the lepton and the Higgs boson pair in the laboratory frame, $y_{lh}$, has same sign. For events with $y_{lh}>0$, we find $A_{\rm CP}^{pp}\simeq -13\%$ and $-3.6\%$ for $\Lambda=500\,$GeV and $1\,$TeV, respectively. Interestingly, using the lepton momentum instead of the $W$ momentum leaves the asymmetry almost intact. Since the  ``lepton-based'' up-down asymmetry avoids reconstructing the $W$ boson rapidity it is likely to be the most effective probe of the CP-odd $hWW$ vertex.
If statistics permits, better sensitivity to larger scales $\Lambda$ can be obtained by cutting harder on the invariant mass of the final states or/and on their average rapidity.
We show in Fig.~\ref{figApp} the expected $A_{\rm CP}$ at the 14$\,$TeV LHC as a function of cuts imposed on $Wh$ and $lh$ systems.

The most important feature that the plots in Fig.~\ref{figApp} reveal is that the up-down asymmetry can be sizable, its magnitude reaching as much as $40\%$-$50\%$ even when the scale $\Lambda$ suppressing the operator contributing to $C_W$ is as large as 1~TeV\@. The top (bottom) plot in Fig.~\ref{figApp} uses as variables the rapidity and invariant mass of the $Wh$ ($lh$) system. The two plots in Fig.~\ref{figApp} are quite similar qualitatively, which shows that the process of reconstructing the momentum of the $W$ is not necessary and the experiments can rely on the straightforward measurement of the charged lepton momentum from $W$ decays. The asymmetry is small near the production threshold that is at small invariant masses of $Wh$, or of $lh$ by proxy, as we already noted in Section~\ref{sec:helicity}. The CP-violating coupling $C_W$ is proportional to momenta, thus it vanishes at threshold. For very large $Wh$ or $lh$ invariant masses  the asymmetry decreases somewhat as the total cross section starts receiving sizable contributions from the square of $C_W$, which is negligible at small invariant masses, and this dampens the magnitude of the asymmetry. This effect is particularly pronounced for smaller scales $\Lambda$. Tightening the cut on the rapidity of $Wh$ or $lh$ yields modest increases of the magnitude of the asymmetry as this cut reduces the probability of misidentifying the quark direction, however tighter cuts swiftly decrease statistics. 

\begin{figure}[htb]
\centering
\includegraphics[scale=0.6]{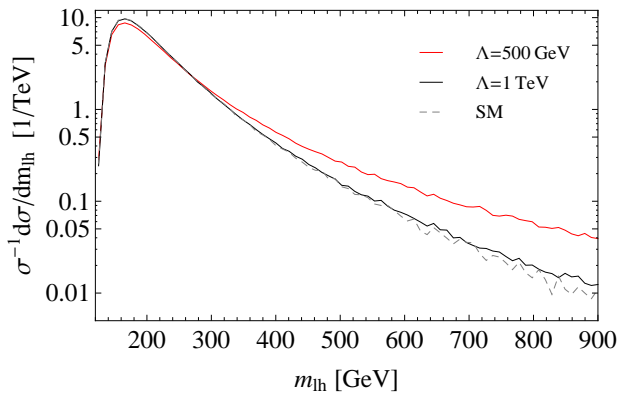}
\caption{Differential distribution of $pp\to W^+h \to l^+\nu bb$ normalized to the total cross-section as a function of the $l^+h$ invariant mass $m_{lh}$.}
\label{figxsdiff}
\end{figure}
Fig.~\ref{figxsdiff} shows the differential cross section for the associated Higgs production in the presence of  the $C_W$ coupling in \Eref{hVV}. The cross section is plotted as a function of the $lh$ invariant mass and illustrates 
how the contribution of $C_W$ to the cross section grows with increasing invariant mass $m_{lh}$. For $\Lambda=500$~GeV, the cross section receives substantial contributions from $C_W$ at large $m_{lh}$ as is expected to arise from the presence of irrelevant operator.  For $\Lambda=1$~TeV, the cross section barely differs from the SM cross section. Despite this, the up-down asymmetry can be large when  $\Lambda=1$~TeV because the asymmetry depends linearly on $C_W$, while the cross section scales as $C_W^2$.
Note that naive dimensional analysis suggests that the scale in which the effective field theory description is expected to break down is roughly $4\pi \Lambda/g$. This is well above
the $Wh$ invariant mass scale that can be probed by the experiments in the near future.

\section{Effective operator basis}
\label{sec:eft}

We now turn to a possible origin of the couplings in \Eref{hVV} and to constraints on these couplings. As of yet there is no sign of physics beyond SM, so it is compelling to assume that new physics is heavy compared to the masses of SM particles and that new physics respects the SM gauge symmetries. At energies too low to produce new states on-shell one can characterize new physics by effective operators involving SM fields only. The operators of dimension six with linearly realized electroweak symmetry were classified in Ref.~\cite{dim6}.

Dominant new physics contributions to the $hV_\mu V_\nu$ vertex in Eq.~\eqref{hVV} then arise from the Lagrangian 
\beq
\mathcal{L}_{d=6} = \sum_i c_i \mathcal{O}_i + \tilde c_i \widetilde{\mathcal{O}}_i\,,
\label{eq:efflag}
\eeq
where $\mathcal{O}_i$ denote the CP-even operators
\begin{gather}
\mathcal{O}_{DH} = H^\dagger H |D_\mu H|^2\,,\ \mathcal{O}_{H}= |H^\dagger D_\mu H|^2\,,\nonumber \\
\mathcal{O}_{WW} = \frac{g^2}{2} H^\dagger H\, W_{\mu\nu}^a W^{\mu\nu\,a}\,,\ 
\mathcal{O}_{BB} = \frac{g^{\prime\,2}}{2} H^\dagger H\, B_{\mu\nu} B^{\mu\nu}\,,\nonumber \\ 
\mathcal{O}_{WB} = gg' H^\dagger \sigma^a H\, W_{\mu\nu}^a B^{\mu\nu}\,, \label{eq:d6even}
\end{gather}
while $\widetilde{\mathcal{O}}_i$ denote the CP-odd operators 
\begin{gather}
\widetilde{\mathcal{O}}_{WW} = \frac{g^2}{2} H^\dagger H\, W_{\mu\nu}^a \widetilde{W}^{\mu\nu\,a}\,,\ 
\widetilde{\mathcal{O}}_{BB} = \frac{g^{\prime\,2}}{2} H^\dagger H\, B_{\mu\nu} \widetilde{B}^{\mu\nu}\,,\nonumber \\ 
\widetilde{\mathcal{O}}_{WB} = gg' H^\dagger \sigma^a H\, W_{\mu\nu}^a \widetilde{B}^{\mu\nu}\,. \label{eq:d6odd}
\end{gather}
We denoted the SM Higgs doublet as $H$, the $SU(2)_L$ and $U(1)_Y$ gauge field strength tensors as $W_{\mu\nu}^a$ and $B_{\mu\nu}$, respectively, and the dual field strengths as $\widetilde{V}_{\mu\nu}=\epsilon_{\mu\nu\alpha\beta}V^{\alpha\beta}/2$. 

The operators $\mathcal{O}_{WB}$ and $\mathcal{O}_{H}$, included in \Eref{eq:d6even}, are in a one-to-one correspondence with the $S$ and $T$ oblique parameters~\cite{peskintakeuchi}. The precision electroweak constraints on $S$ and $T$ are so stringent that the coefficients of these operators $c_{WB}^{-1/2},c_{H}^{-1/2}\gtrsim \mathcal{O}(8\,$TeV$)$~\cite{ewpt}. For the purpose of our discussion, we can assume that $c_{WB}$ and $c_H$ are negligibly small as direct measurements of Higgs couplings do not have enough accuracy to probe such high scales in the foreseeable future. The remaining operators in \Eref{eq:d6even} are not constrained  by the LEP experiments as they reduce to SM gauge kinetic terms when the Higgs doublet is substituted by its vacuum expectation value (vev) $v$. Likewise, the CP-odd operators in \Eref{eq:d6even} give boundary terms when the Higgs doublet is substituted by its vev. These operators contribute in perturbation theory only when the physical Higgs scalar, from expanding $H^\dagger H=\frac{v^2}{2} + v h + \frac{h^2}{2}$,  is involved in the interaction. Hence, there are no direct LEP bounds on these operators. 

In terms of the CP-odd operator coefficients defined above, we obtain the following CP-violating couplings for the $hWW$ and $hZZ$ defined in \Eref{hVV}
\begin{gather}
C_W= 4 \tilde{c}_{WW}\,,\nonumber  \\
 C_Z= 4 (\tilde{c}_{WW} c_w^4 + \tilde{c}_{BB} s_w^4 + 2 \tilde{c}_{WB} s_w^2 c_w^2)\,.  \label{eq:ffd6}
\end{gather}
For completeness we also give the CP-preserving couplings in \Eref{hVV} in terms of the CP-even operator coefficients in \Eref{eq:efflag}
\beq
B_W=  4 c_{WW}\,,  \quad B_Z= 4 (c_{WW} c_w^4 +  c_{BB} s_w^4)\,,
\eeq
and~\footnote{In the presence of $\mathcal{O}_{DH}$, the Higgs vev is related to the Fermi constant $G_F$ through $v= (\sqrt{2}G_F)^{-1/2}\times\left[1-c_{DH}(\sqrt{2}G_F)^{-1}/4 + \mathcal{O}(c_{DH}^2)\right]$.} 
\beq
A_V=1 +c_{DH}\frac{v^2}{2} - p_1 \cdot p_2 \, B_V + \mathcal{O}(c^2_{DH})\,,
\eeq
where $s_w$ and $c_w$ are the sine and cosine of the weak mixing angle, respectively. Assuming that new physics is heavy, the leading contributions to the form-factors $B_V$ and $C_V$ computed above are momentum independent. The factor of $1$ in $A_{V}$ denotes the SM contributions and does not originate from higher-dimensional operators in \Eref{eq:efflag}. Note that non-standard $A_V$ and $B_V$ couplings always lead to subdominant $\mathcal{O}(c^2)$ effects on the up-down asymmetry.

It is worth pointing out that the $hWW$  and $hZZ$ vertices listed in \Eref{eq:ffd6} receive contributions from different effective operators and are not always simply proportional to each other. At first sight, this might suggest a large violation of custodial symmetry, but in fact violation of custodial symmetry is only by the gauging of the hypercharge which is the same type of custodial symmetry breaking that is already present in the SM\@. Of the six operators that contribute to  \Eref{eq:ffd6} two, $\mathcal{O}_{WW}$ and $\widetilde{\mathcal{O}}_{WW}$,  preserve custodial symmetry. The remaining four operators, including  $\mathcal{O}_{DH}$, violate custodial symmetry when $g'\neq0$. The operator $\mathcal{O}_{DH}$ is proportional to the Higgs kinetic energy in the SM and gives identical contributions to $A_W$ and $A_Z$, which is an artifact of SM normalization in \Eref{hVV}, when in fact the $hWW$ and $hZZ$ couplings are different. One way of understanding that the custodial symmetry is broken according to the same pattern by the operators $\mathcal{O}_{DH}$, $\mathcal{O}_{BB}$,  $\widetilde{\mathcal{O}}_{BB}$, and $\widetilde{\mathcal{O}}_{WB}$ is by gauging the full $SU(2)_R$ symmetry. Under that gauging the custodial symmetry is restored and there are two triplets of vector bosons with couplings that respect the diagonal custodial symmetry. This can be contrasted with custodial symmetry breaking by the operator $\mathcal{O}_{H}$ which persists even in the limit $g'\to 0$. Since the differences between the $hWW$ and $hZZ$ couplings are caused by the hypercharge only, potential discrepancies in these couplings can be natural and do not require new sources of custodial symmetry violation. 

Comparing the expressions for $C_W$ and $C_Z$ in \Eref{eq:ffd6} illustrates why it is worth measuring $C_W$ even if $C_Z$ could be constrained to be small. These CP-violating couplings arise from independent operators and probe different linear combinations of their coefficients. It is likely that $C_Z$ can either be measured or tightly bound using the $h\to4 l$ channel (though only for the case where the $Z Z^*$ invariant mass is equal to the Higgs mass). In fact, one other linear combination of these operators is already bounded by constraints on the electric dipole moments (EDM) of the electron. The CP-violating coupling of the Higgs to two photons contributes to the electron EDM assuming that Higgs coupling to electrons is SM-like~\cite{Pospelov}. Electron EDM predominantly\footnote{The electron EDM receives contributions from CP violating parts of both $h\gamma \gamma$ and $h Z\gamma$, but the contribution of  $h Z\gamma$ is suppressed by the small value of $1- 4 s_w^2$.} restricts the operator $\frac{e^2}{2} H^\dagger H F_{\mu\nu}\widetilde{F}^{\mu\nu}$, with $e$ the electric charge and $F_{\mu\nu}$ the photon field strength. The coefficient  of this operator, $\tilde{c}_{\gamma\gamma}= \tilde{c}_{WW}  + \tilde{c}_{BB}  -2  \tilde{c}_{WB}$, is linearly independent of the expressions for $C_W$ and $C_Z$. The bounds from electron EDM, $\tilde{c}_{\gamma\gamma}^{-1/2}\gtrsim \mathcal{O}(7\,$TeV$)$~\cite{Pospelov}, make probing CP violation in the $h\to \gamma \gamma$ decays particularly challenging, for example using the method suggested in Ref.~\cite{Voloshin}.

\section{Conclusions}
\label{sec:conclusions}
We proposed a new method of measuring CP-violating couplings of Higgs to $W$ bosons using associated Higgs production. Our observable is based on counting the number of leptons produced in $W$ decays with momenta above or below the plane containing the beam and Higgs momentum. The orientation of that plane is established by the cross product of quark and Higgs momenta.  We showed that our observable is quite a sensitive probe of CP-violating $hWW$ coupling. We demonstrated the measurement of the asymmetry can be done at 14~TeV LHC using rapidity cuts to select quark direction. Our observable is very straightforward to implement experimentally once Higgs boson is reconstructed and the associated $W$ boson is selected through cuts on the lepton momentum and missing energy. The main obstacle is low statistics due to small production cross section and reconstruction efficiencies.  

Disentangling the nature of Higgs couplings to other SM particles is a crucial next step for either confirming or disproving the validity of the SM at yet un-probed energy scales. Higgs couplings could simply differ in magnitudes from those predicted by the SM, but they could also contain terms of different symmetry properties. In particular, CP violation in the Higgs sector is not as tightly constrained as it is for various interactions involving light SM fermions. While CP violation in the $hZZ$ interaction vertex can be tested relatively easily using the $h\to 4l$ channel, the $hWW$ vertex is more difficult to probe and yet it may contain independent information about CP violation in the Higgs sector. 

We have not performed any detailed studies of experimental intricacies such as detector resolution, acceptance or pileup effects. Despite being crucial for optimizing the cuts, simulating these effects carefully is beyond the scope of this work. We expect however that the numerical value of the asymmetry will clearly be somewhat reduced by these experimental effects  compared to our predictions. Nonetheless, due to the simplicity of the proposed up-down asymmetry we expect it will still be a useful observable to measure irrespectively of experimental challenges.

\section*{Acknowledgements}
 We thank Gavin Salam, Martin Schmaltz, Andrea Wulzer for discussions, and Maxim Pospelov for helpful correspondence. The work of GP is supported by grants from GIF, ISF, Minerva and the Gruber award. The work of HS is supported by a grant from CNPq/CsF program \#202129/2012-8. WS thanks the CERN Theory group for its hospitality and the US Department of Energy and the Simons foundation for their support. This work was partially supported by a grant from the Simons Foundation (\#267592 to Witold Skiba).


\begin{thebibliography}{plain}

\bibitem{ATLASdisco}
 G.~Aad {\it et al.}  [ATLAS Collaboration],
  Phys.\ Lett.\ B {\bf 716}, 1 (2012)
  [arXiv:1207.7214 [hep-ex]].

\bibitem{CMSdisco}
  S.~Chatrchyan {\it et al.}  [CMS Collaboration],
  Phys.\ Lett.\ B {\bf 716}, 30 (2012)
  [arXiv:1207.7235 [hep-ex]].

\bibitem{Higgsfits}
A.~Falkowski, F.~Riva and A.~Urbano,
  arXiv:1303.1812 [hep-ph];
 P.~P.~Giardino, K.~Kannike, I.~Masina, M.~Raidal and A.~Strumia,
  arXiv:1303.3570 [hep-ph].
  
\bibitem{ManoharWise}
A.~V.~Manohar and M.~B.~Wise,
  Phys.\ Lett.\ B {\bf 636}, 107 (2006)
  [hep-ph/0601212].
  
 \bibitem{assocprod}
   G.~Isidori and M.~Trott,
  arXiv:1307.4051 [hep-ph].

\bibitem{CMS_4leptons}
  S.~Chatrchyan {\it et al.}  [CMS Collaboration], CMS-PAS-HIG-13-002.

\bibitem{ATLAS_4l}
  G.~Aad {\it et al.}   [ATLAS Collaboration], ATLAS-CONF-2013-013  

\bibitem{Godbole} 
  R.~Godbole, D.~J.~Miller, K.~Mohan and C.~D.~White,
  arXiv:1306.2573 [hep-ph].

\bibitem{Zeppenfeld} 
T.~Plehn, D.~L.~Rainwater and D.~Zeppenfeld,
  Phys.\ Rev.\ Lett.\  {\bf 88}, 051801 (2002)
  [hep-ph/0105325];
  V.~Hankele, G.~Klamke, D.~Zeppenfeld and T.~Figy,
  Phys.\ Rev.\ D {\bf 74}, 095001 (2006)
  [hep-ph/0609075].

\bibitem{Gao} 
  Y.~Gao, A.~V.~Gritsan, Z.~Guo, K.~Melnikov, M.~Schulze and N.~V.~Tran,
  Phys.\ Rev.\ D {\bf 81}, 075022 (2010)
  [arXiv:1001.3396 [hep-ph]].

\bibitem{CTEQ}
J.~Pumplin, D.~R.~Stump, J.~Huston, H.~L.~Lai, P.~M.~Nadolsky and W.~K.~Tung,
  JHEP {\bf 0207}, 012 (2002)
  [hep-ph/0201195].
  
\bibitem{MG5}
J.~Alwall, M.~Herquet, F.~Maltoni, O.~Mattelaer and T.~Stelzer,
  JHEP {\bf 1106}, 128 (2011)
  [arXiv:1106.0522 [hep-ph]].
  
\bibitem{dim6}
W.~Buchmuller and D.~Wyler,
  Nucl.\ Phys.\ B {\bf 268}, 621 (1986);
  B.~Grzadkowski, M.~Iskrzynski, M.~Misiak and J.~Rosiek,
  JHEP {\bf 1010}, 085 (2010)
  [arXiv:1008.4884 [hep-ph]].


\bibitem{peskintakeuchi}
M.~E.~Peskin and T.~Takeuchi,
  Phys.\ Rev.\ Lett.\  {\bf 65}, 964 (1990);
 M.~E.~Peskin and T.~Takeuchi,
  Phys.\ Rev.\ D {\bf 46}, 381 (1992).

\bibitem{ewpt}
 R.~Barbieri and A.~Strumia,
  Phys.\ Lett.\ B {\bf 462}, 144 (1999)
  [hep-ph/9905281];
   R.~Barbieri, A.~Pomarol, R.~Rattazzi and A.~Strumia,
  Nucl.\ Phys.\ B {\bf 703}, 127 (2004)
  [hep-ph/0405040];
  Z.~Han and W.~Skiba,
  Phys.\ Rev.\ D {\bf 71}, 075009 (2005)
  [hep-ph/0412166].
  
\bibitem{Pospelov}
D.~McKeen, M.~Pospelov and A.~Ritz,
  Phys.\ Rev.\ D {\bf 86}, 113004 (2012)
  [arXiv:1208.4597 [hep-ph]].
  
\bibitem{Voloshin}
 M.~B.~Voloshin,
  Phys.\ Rev.\ D {\bf 86}, 093016 (2012)
  [arXiv:1208.4303 [hep-ph]].


\end{thebibliography}
\end{document}